# BISCAY: Practical Radio KPI Driven Congestion Control for Mobile Networks

Jon Larrea, Tanya Shreedhar and Mahesh K. Marina

## ABSTRACT

Mobile application performance relies heavily on the congestion control design of the underlying transport, which is typically bottlenecked by cellular link and has to cope with rapid cellular link bandwidth fluctuations. We observe that radio KPI measurements from the mobile device chipset can be exploited for precise and timely measurement of available bandwidth on the cellular link. Building on this insight, we propose BISCAY, a practical and radio KPI-driven congestion control system design for mobile networks. BISCAY leverages OPENDIAG, the in-kernel real-time radio KPI extraction tool we introduce in this paper, along with our KPI-based accurate bandwidth determination layer towards dynamically adjusting the congestion window to optimally use the available bandwidth while keeping delay to the minimum. Our solution is practical and deployable, as shown through our implementation of BISCAY and OPENDIAG on unrooted Android 5G phones. We extensively evaluate BISCAY against different state-of-the-art congestion control designs including BBR and CUBIC with emulations driven by real measurement traces as well as real-world experiments spanning diverse 4G and 5G scenarios, and show that it provides significant average and tail delay improvements (typically over 90% reduction) while yielding better or similar throughput. These gains are enabled by 100× improvement in the granularity of on-device radio KPI measurements with OPENDIAG compared to existing alternatives like MobileInsight.

## 1 INTRODUCTION

Mobile cellular networks enable ubiquitous and on-the-move connectivity for end devices. Global mobile subscriptions already exceed 8 billion (current world population), significantly dwarfing fixed broadband users and rapidly evolving from being 4G centric today to becoming primarily 5G based in the coming few years [39]. Despite the rapid rollout and adoption of 5G, several measurement studies show that it is yet to fully deliver on high throughput and low delay required to support many applications [44, 54, 55]. While traditional mobile applications have predominantly relied on downstream traffic to deliver content to users, there is an increasing number of next-generation applications and use cases that generate significant uplink traffic [54]. Applications such as cloud gaming, augmented reality (AR), virtual reality (VR), video conferencing, backup services and high-definition video streaming (e.g., live broadcasting) demand substantial uplink capacity to transmit data in real time. These uplink-focused applications highlight the growing importance of addressing uplink traffic to ensure consistent performance and user experience in modern mobile networks.

Our focus in this paper is on high-performance transport protocols for mobile networks and specifically on improving congestion control (CC). CC plays a pivotal role in influencing the performance of a wide variety of applications, including video streaming [42, 62, 63], real-time analytics

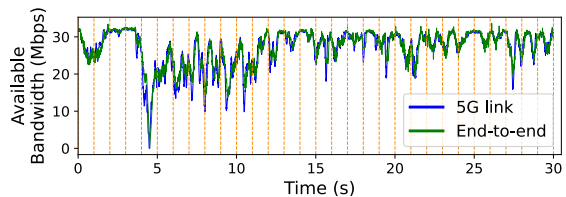

**Figure 1:** Measured trace of end-to-end and 5G link available bandwidth fluctuations while moving with a commodity 5G phone.

[107], V2X communications [74] or AR/VR services [60, 68]. The figure of merit for effective congestion control (CC) is to maximize delivered throughput while minimizing delay, captured together through the combined power (throughput/delay) metric. Optimizing this metric in turn requires matching traffic sent into the network with the bandwidth-delay product (BDP) of the network path as reflected in the rule of thumb "keep the pipe just full, but no fuller" [58].

Tracking the above-mentioned optimal operating point for CC in mobile networks is hard for two reasons: (1) Cellular link is typically the bottleneck, as was also observed in previous works [63, 102, 104, 106, 111]; (2) the available bandwidth for a mobile device over the cellular link exhibits frequent fluctuations due to the inherent nature of the wireless channel, user contention, mobility, and handovers, as also noted in prior work [46, 63, 102, 106, 111]. Figure 1 illustrates both these challenges using a measurement trace from our dataset, collected with a commodity phone connected to 5G network while moving (more details in §5.3.1). This highlights *the importance of having precise and timely knowledge of available bandwidth on the mobile network segment of the path to design effective CC.* Prior works focusing on CC for mobile networks (discussed in detail in §2.1) broadly fall into two categories: (i) those that rely on estimation or prediction of the available bandwidth in various ways (e.g., Sprout [102], PROTEUS [109], Verus [111], PropRate [65], ExLL [79]); (ii) the works that require support from the network infrastructure like base stations (e.g., ABC [46], DChannel [86]) or additional hardware like cellular sniffers (e.g., PBE-CC [106]). The former set of works are estimates and thus inherently limited in their accuracy, whereas the latter category comes with deployability limitations and challenges.

Our key insight in this paper is that direct measurement of available bandwidth on cellular link is better than estimation (end-to-end or otherwise), and that routinely and implicitly measured key performance indicators (KPIs) in the device radio chipset can be leveraged towards this available bandwidth measurement at no additional probing cost. Simple as it may seem, exploiting this insight however presents several challenges: (1) which among the numerous (thousands of) radio KPIs are relevant for available bandwidth determination and how can they be used to compute it?; (2) how to extract the typically inaccessible radio KPIs of interest efficiently and in real-time?; (3) how to leverage the fine-grained radio KPI driven available bandwidth measurement of the cellular link to effectively drive congestion window adaptation?; (4) finally, there is the cross-cutting challenge of answering the above questions in a practical and easily deployable manner.

| CCA | Tput | Avg Delay | Tail Delay | CCA | Tput | Avg Delay | Tail Delay |
|-----|------|-----------|------------|-----|------|-----------|------------|
| BBR | 1.03× | 58.51% | 41.18% | PCC | 1.84× | 94.97% | 95.01% |
| CUBIC | 0.96× | 98.74% | 99.03% | Sprout | 1.01× | 80.69% | 71.59% |
| Copa | 1.01× | 96.41% | 96.48% | Verus | 1.56× | 90.83% | 97.28% |
| LEDBAT | 1.0× | 92.9% | 88.55% | Vivace | 1.72× | 93.03% | 94.05% |

**Table 1:** Summary of performance gains with BISCAY (in terms of throughput increase factor and percentage of average/tail latency reduction) relative to existing CCAs.

We propose BISCAY, a practical and radio KPI driven congestion control system design that leverages the above insight and addresses the associated challenges. Through a deeper look at the 3GPP standards [16], we identify the key set of KPIs (e.g., transport block size, number of physical resource blocks) that allow a device to determine the current available bandwidth over the cellular link. At the core of BISCAY is an in-kernel tool termed OPENDIAG, designed for real-time extraction of any specified set of KPIs from the chipset using the integrated Diag interface. This enables accurate computation of available bandwidth at fine time granularity. BISCAY CC mechanism uses the cellular link bandwidth measurement via OPENDIAG to set the congestion window. In the scenario where the bottleneck lies within the wired segment, which is quite uncommon, BISCAY leverages existing end-to-end bandwidth estimation methods.

We implement BISCAY and OPENDIAG for Android devices. We also provide a well-defined API for user space transport protocol implementations to leverage radio KPIs through a library called *libOD*. Notably, unlike other existing tools for radio KPI extraction (e.g., MobileInsight [66, 67, 97]), OPEN-DIAG enables two orders of magnitude finer timescale KPI extraction and also does not require rooting the device. The above combined with the fact that BISCAY is a device-centric solution makes it highly practical and readily deployable. Upon publication of this work, we intend to make BISCAY and OPENDIAG available, along with our companion tools and measurement traces used for evaluations, to benefit the research community.

We conduct extensive experimental evaluations of BISCAY using its above outlined implementation in comparison with a wide range of existing congestion control algorithms (CCAs). We do this in two ways: (1) using the Pantheon emulator [110], driven by a large number of measurement traces spanning diverse scenarios we collected using commodity 5G phones augmented with OPENDIAG. Each trace consists of backlogged UDP throughput and companion radio KPI measurements. (2) real-world experimentation of BISCAY in the wild (private and public networks) using commodity 5G phones in comparison with BBR [34] and CUBIC [49].

Table 1 summarizes some of the key results (elaborated in §5). Overall, BISCAY yields significant reductions in average and tail delays (at least 50% but typically over 90% reduction) compared to all existing CCAs, while delivering better or similar throughput. We also perform detailed evaluations of OPENDIAG relative to existing radio KPI extraction approaches including MobileInsight [97] and Android Telephony API [45], and show that OPENDIAG provides a 100× improvement in the granularity of KPI measurements with respect to these alternatives.

In summary, we make the following key contributions:

- We highlight an unexplored opportunity for effective CC over mobile networks by leveraging radio KPI measurements *from the mobile device chipset* and introduce BISCAY, a new device-centric cellular CC system design that fully exploits it (§3).
- As an enabler for BISCAY and other fine-grained radio KPI data driven use cases, we develop OPENDIAG (§3.2.3), the first real-time radio KPI extraction tool for commodity mobile devices that does not need rooting and allows arbitrary set of radio KPIs to be obtained from the radio modem at 10*ms* time granularity – 100× improvement over common alternatives like MobileInsight [97] and Android Telephony API [45].
- Through extensive evaluations of BISCAY and OPENDIAG (§5) with their respective in-kernel implementations on a commodity Android 5G phone (§4) and using our collected 5G performance measurement traces as well as real-world experimentation, we not only demonstrate the practicality of the proposed BISCAY cellular CC approach but also its effectiveness relative to existing CCAs in significantly reducing the average and tail delays while delivering better or similar throughput.

## 2 BACKGROUND AND RELATED WORK

Here we focus on discussing related works on cellular congestion control and on-device mobile network monitoring tools. The appendix A.1 provides the relevant background on the 5G networking stack on mobile devices while A.2 gives an overview of available on device channels for communication with the radio modem.

### 2.1 Congestion Control Mechanisms

**End-to-end approaches.** Traditional loss-based CCAs like NewReno [48] and CUBIC [49] reduce congestion window (cwnd) based on packet loss, but they react slowly to cellular network fluctuations, leading to delays. Delay-based alternatives such as Vegas [31], FastTCP [100], Copa [29] and LEDBAT [87] which rely on round-trip time (RTT) for congestion management are also unable to react to the cellular link fluctuations given that they operate at RTT granularities making them unable to capture the fine grain bandwidth fluctuations experienced in the air interface leading to inefficient resource utilization. More recent hybrid approaches (e.g., BBR [34], TCP-Illinois [69], Compound TCP [94]) rely on a combination of signals to adjust the congestion window. While these schemes considerably improve bandwidth utilization, they have other issues such as RTT unfairness, TCP unfriendliness and robustness. Several learning-based CCAs have also been proposed recently (e.g., PCP [26], PCC [36], PCC-Vivace [37], Remy [101], Indigo [110], Orca [23]) as a way to adapt to varying network conditions. But they have a huge associated cost for training and limited generalization, as a model trained on a specific network cannot be used in a different network.

**Cellular oriented approaches.** Several congestion control approaches have been proposed in the recent past specifically targeting cellular networks (e.g., Sprout [102], PRO-TEUS [109], Verus [111], PropRate [65], ExLL [79], ABC [46], PBE-CC [106], DChannel [86]). Earlier methods in this category treat the cellular link as a black box but differ in how they estimate or predict its dynamics. For instance,

Sprout [102] and PROTEUS [109] employ short-term forecasting, PropRate [65] uses continuous probing. Verus [111] uses a BBR-like hybrid approach with delay measurements designed for cellular links, and ExLL [79] monitors packet arrival patterns at the cellular receiver. Recently mBBR [112], an enhancement over BBR was proposed which addresses the challenge of high packet loss rates and bandwidth costs for rate-limited mobile networks. Works such as LDRP [96] aims for uplink latency reduction through application-layer latency estimation using dummy packets. This redundant mechanism, which is already done by transport protocols (e.g., BBR) without introducing *dummy* probing packets in the network, results in network and energy overheads.

More recent works assume network infrastructure or specialized hardware support. Some, resembling router-based or Active Queue Management (AQM) based strategies like CoDel [75], such as CQIC [70], ABC [46, 47], XRC [56] and DChannel [86], rely on cross-layer cellular base station information. These solutions demand extensions to both the base station and mobile device, presenting deployment issues. Other methods leverage information from cellular sniffers (e.g., [3, 33, 41, 59, 92, 105]) for optimized congestion control, as shown in piStream [103] and PBE-CC [106]. PBE-CC [106], a representative example, introduces a hardware-assisted CCA that leverages fine-grained low-layer data from multiple sniffers. However, its dependence on external hardware (programmable SDRs) and functionalities beyond 3GPP specifications [15, 16, 21] hinders its deployment on standard mobile devices. Moreover, the energy-efficient nature of modern devices fundamentally limits its deployment given that PBE-CC relies on brute force and exhaustive tree searches to recover information from the wireless link [105].

Compared to the existing solutions, 1) BISCAY can adapt to the frequent and unpredictable fluctuations of the wireless bandwidth unlike end-to-end approaches which rely on RTT, resulting in fewer queue buildups and thus lower delays; 2) BISCAY provides significantly more accurate bottleneck bandwidth estimate compared with the cellular-specific solutions that treat the air interface as a black box, resulting in throughput maximization while keeping the delays at a minimum; and 3) BISCAY's device-centric design eliminates the need for support from the routers on the path or external devices (such as sniffers), facilitating its deployment and adoption.

## 2.2  On-Device Mobile Network Monitoring

Most commercial on-device mobile network monitoring tools function as trace-collector tools (e.g., [24, 52, 53, 82]) and are designed for offline analysis, where the data processing happens on a separate machine. Alternatively, commercial tools capable of *online monitoring* [80, 84, 90] are limited to on-device visualization of the data for service quality assessment or RF troubleshooting during field testing. However, they lack the ability to forward the collected data to another on-device consumer in real-time. In the open-source realm, some tools (e.g., [30, 76, 78]) dump in near-real-time basic information from some of the mobile layers for use on an external machine connected to the phone.

Among the on-device radio KPI extraction solutions, alternatives such as the Telephony API [45] or the minimal driver prototypes of the Qualcomm Diag protocol [77] (e.g., [28, 43,

72, 91]) limit the range and granularity of available KPIs (also shown in our evaluations §5). MobileInsight [66, 67, 97] is an open-source tool for on-device radio KPI extraction. While it offers the potential to retrieve all chipset-measured KPIs, it requires the device to be rooted, which introduces multiple vulnerabilities [95] and hampers widespread deployment. Furthermore, its online KPI extraction suffers from coarse (1s) data granularity due to its original offline design [4], necessitating user extensions to its mobile app [5] to be able to forward KPI data to other on-device consumers. Some research studies like CLAW [104] and PERCEIVE [63] have leveraged MobileInsight for applications such as optimizing web page latency and uplink throughput prediction for client-side rate adaption in video streaming, respectively. However, these solutions face challenges due to MobileInsight's coarse granularity, as highlighted in §5. In contrast, our solution OPENDIAG facilitates fine-grained on-device and online radio KPI extraction, crucial for effective cellular CC, and does not require device rooting.

## 3  DESIGN

We aim to utilize untapped potential by leveraging ongoing KPI measurements from device radio chipsets for timely determination of available cellular link bandwidth, thus enhancing cellular congestion control. We begin by discussing the challenges. We then present an outline of our congestion control system, BISCAY, followed by a description of its components, including the real-time radio KPI extraction tool called OPENDIAG.

### 3.1  Challenges

**Cellular link bandwidth determination.** The mobile device's 4G/5G radio chipset automatically captures a multitude of KPIs with millisecond-level granularity. Among these, certain KPIs like Channel Quality Indicator (CQI) are vital for mobile network and base station functions (e.g., MAC resource scheduling), while others aid in device-side monitoring and diagnostics (e.g., radio measurements for drive test minimization [22]). Our primary challenge is to identify the specific subset of radio KPI measurements from the device chipset that are important for calculating cellular link capacity and available bandwidth.

**Real-time radio KPI extraction.** In the context of this work, where available cellular link bandwidth is derived from low-level radio KPIs on devices, the age (time since measurement) and granularity (measurement frequency) of these KPIs significantly impact the precision of current bandwidth estimation. Fresh and finely-grained measurements are critical for accuracy. However, as discussed in §2.2, existing on-device radio KPI monitoring tools do not meet this requirement. Our second challenge is to solve this.

**On-device radio KPI based congestion control.** Apart from the above two challenges, we need to identify how and when to use the cellular link bandwidth measurement information from the congestion control perspective. Particularly when the cellular link is the bottleneck (a common scenario), the challenge is how to incorporate the measured cellular link bandwidth value for congestion control across multiple concurrently active flows.

**Deployability.** The ease of deployment of a congestion control (CC) design plays a key role in its widespread acceptance and adoption. Currently, the only CC design that leverages real-time radio KPIs requires bulky external hardware in the form of cellular sniffers [106] plugged into a phone, rendering it impractical for deployment. Ideally, both the CC system design and the radio KPI extraction framework should seamlessly operate on standard devices, without burdening users or necessitating device rooting.

## 3.2 BISCAY

### 3.2.1 System Overview. Figure 2 gives an overview of BISCAY, our proposed CC system design. OpenDIAG, our *real-time radio KPI extraction layer* interfaces with the Diag module to collect radio KPIs across various layers of the mobile network stack. OpenDIAG is accessed by BISCAY's *cellular link bandwidth determination layer* to obtain the current available bandwidth for the cellular link. This information, along with end-to-end bandwidth estimation (obtained by leveraging existing techniques), is sent to BISCAY's *bottleneck determination layer* to determine connection's bottleneck bandwidth. The result is combined with the RTT estimation (derived from the Linux TCP stack machinery) to obtain an adjusted congestion window value which is forwarded to the kernel's TCP mechanism. All layers, except OpenDIAG, collectively form the BISCAY's CC module.

### 3.2.2 Cellular link bandwidth determination layer. Unlike prior approaches [63, 88, 89, 104, 106] that simply utilize raw radio KPIs with correlation and prediction models for bandwidth estimation, we argue (as shown in §5) that replicating the modem's internal method, as outlined by 3GPP for both 4G [21] and 5G [1], to calculate the available bandwidth offers a more accurate and robust approach to determine cellular link bandwidth. Our focus centers on 5G bandwidth determination, which is similar to 4G.

The 3GPP specification TS 38.306 - 4.1.2 [1] defines a formula for calculating the maximum ideal throughput of a device (UE in 3GPP terminology), considering input parameters like aggregated component carriers, maximum number of MIMO layers, modulation order, and resource block allocation. However, this formula does not accurately represent the current available bandwidth; it estimates achievable throughput under ideal conditions.

In contrast, another 3GPP specification [16] outlines how UEs calculate their available uplink bandwidth using Transport Block Size (TBS) determination method, based on grants from base stations via Downlink Control Information (DCI) messages. The modem uses it to calculate the uplink bandwidth. This approach has two phases in 5G: TBS index calculation (phase 1) and bandwidth calculation (phase 2). In 5G, the TBS index is dynamically calculated per Transmission Time Interval (TTI) (as per TS 38.214 - 5.1.3 [16]), using parameters like modulation order and coding rate derived from Modulation and Coding Scheme (MCS), redundancy version or the number of scheduled OFDM symbols specified by the base station, based on UE-reported measurements. In 4G, the TBS index is pre-calculated and can be obtained using the MCS index. Once there is a TBS index, it is used along with the number of Physical Resource Blocks (PRBs) to determine the bandwidth (multiplied by the number of antennas in case of using MIMO) using pre-defined tables. This process is repeated for each carrier in carrier aggregation (CA). In practice, this process can be simplified as the DCI message decoded via the diagnostic channel already includes the TBS along with the number of PRBs, effectively removing phase 1 of the algorithm. Equation 1 shows the cellular link bandwidth calculation, repeated for every TTI.

$$bw = \sum_{c=1}^{Carriers} (tputTable[PRB(c), TBS(c)] * numAntennas)$$
(1)

Alternatively, the MAC layer generates a diagnostic message summary of the grants received (in bytes) and its utilization. It accounts for carriers, MIMO, and standards (4G, 4G+, 5G NSA/SA), albeit at a coarser 100ms granularity. We evaluate both the granted bytes KPI and the described bandwidth calculation procedure from the CC perspective in §5.

Note that the same DCI messages that carry the uplink grants also carry downlink related grants as defined by 3GPP. These downlink grants contain the TBS and the number of PRBs granted by the base station to the UE for the downlink transmission which can be combined, using Equation 1 and replacing the uplink lookup table by the corresponding 3GPP downlink pre-calculated table [1], to obtain the available bandwidth in the downlink direction.

### 3.2.3 OpenDIAG: Real-time radio KPI extraction layer. Access to KPI metrics immediately after they are generated in the chipset is required to obtain the current cellular link bandwidth in an online manner. However, existing KPI collection tool designs (represented by MobileInsight [66] architecture in Figure 3) lack this functionality as they were not designed with real-time capability in mind. Specifically, these tools suffer from three inherent design limitations that hinder their ability to deliver real-time data.

- *Inter-process communication.* MobileInsight is a user-space application that is made up of two processes communicating via a pipe: diag_revealer (a C application responsible for the message collection) and MobileInsight App (a Java application with a Python interpreter on top responsible for the message processing). The data (diag messages containing KPIs) initially traverse the Diag module which interfaces to the modem diagnostic channel and acts as a data forwarder. In addition to these three entities that are part of the processing chain, MobileInsight must be extended in order to forward the processed KPIs to a consumer application (CC in our case), creating another step in the chain.
- *Processing time.* MobileInsight's packet processing framework parses and extracts a wide range of KPIs from packets. This can vary from a few tens of KPIs in small packets to several hundred or even a thousand KPIs in larger ones. However, for specific applications like ours, this design becomes inefficient, as it necessitates waiting for the processing of all KPIs in selected packets before retrieving the required ones. This is particularly problematic when only a handful of KPIs are necessary.
- *Message granularity.* MobileInsight employs the Diag module to receive packets from the modem. However, this approach lacks insight into the modem's internal workings. Notably, it does not account for the modem's aggressive buffering mechanism, resulting in the release of message batches roughly every second, as demonstrated in §5.

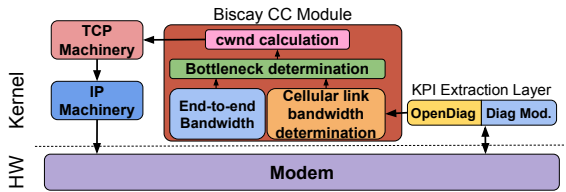

Figure 2: BISCAY congestion control system overview.

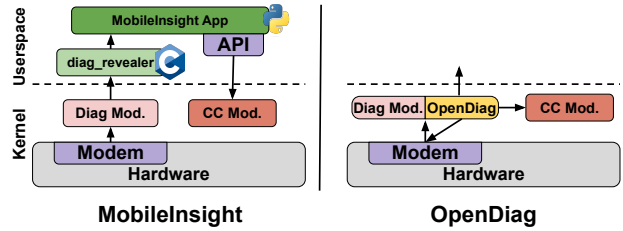

Figure 3: MobileInsight vs OPENDIAG architecture.

To address the above limitations, we present OPENDIAG, an innovative architecture for KPI extraction that enables real-time retrieval of radio KPIs. A comparison of OPENDIAG architecture is shown in Figure 3 alongside the MobileInsight architecture. OPENDIAG provides the following:

*Tightly Coupled Architecture.* In order to minimize the inter-process communication, OPENDIAG tightly integrates with the Diag module (and its equivalent in modern Android versions). This design choice not only eliminates intermediaries like MobileInsight's diag_revealer but also grants access to private Diag module functionality (through exported symbols and bypassing non-essential features) that cannot be accessed as a user. This compact architecture design mandates that OPENDIAG must run within the kernel. OPENDIAG is shipped as a kernel module, so the kernel needs no modifications. This in-kernel design avoids the data from crossing the kernel-user boundary twice before reaching the CC module, generally deployed as a kernel module as well. Notably, eBPF-based strategies [2] prove ineffective in this context, the diagnosis logic resides in the modem and eBPF codelets cannot hook into external hardware.

*Efficient Processing Framework.* Unlike in MobileInsight, where to retrieve one KPI, all the KPIs within that packet have to be processed due to a packet-focused processing framework, OPENDIAG employs a KPI-focused strategy in its processing framework. This approach stipulates that the smallest parsing unit is a single KPI (in contrast to a full packet), enabling the parsing of individual KPIs. The processing efficiency of our KPI-focused methodology is detailed in §5.

*Establishing Control Channel.* Due to its privileged access to Diag module's internal functionality (through the in-kernel architecture), OPENDIAG is able to reach and manipulate modem internals inaccessible from user space. This establishes a control channel with the modem, serving the purpose of coordinating when to drain the internal buffer where the diagnosis messages get queued, effectively removing a key root cause of coarse granularity experienced by prior designs. Moreover, this control channel directs the chipset to offload the minimal necessary packet subset for extracting the user-defined KPIs.

### 3.2.4 Radio KPI based congestion control.
Accurately estimating the bottleneck bandwidth is one of the main challenges of today's CCA. To this end, we present BISCAY, a radio-KPI based TCP CCA that can precisely determine the bandwidth of the cellular link in real time by leveraging cross-layer information. From a design point of view, BISCAY has two main components: a *KPI extraction layer* that is realized through OPENDIAG and the *CC layer* that integrates with the TCP machinery as shown in Figure 2. The KPI extraction layer (OPENDIAG) gathers the required KPIs in real time and makes them accessible within the CC layer.

On the other hand, the CC layer is responsible for converting the extracted KPIs into a throughput value (bits per second) as described earlier in §3.2.2 and then combining the result with an estimated RTT to generate an optimal congestion window *cwnd* that is set into the TCP sending machinery.

In addition, BISCAY is capable of detecting where the bottleneck is located to optimally adapt the congestion window to different scenarios (bottleneck located in the wireless link or in the wired segment). Bottleneck localization in BISCAY works as follows. BISCAY continually calculates both end-to-end bandwidth estimate (achieved by leveraging any of the existing end-to-end bandwidth techniques deployed in other CCAs such as packet sending rate, ACK-based bandwidth estimation, or loss-based estimation) and wireless link bandwidth via the KPI extraction layer. In the typical case where the cellular link is the bottleneck, cellular link bandwidth calculated through the KPIs will be lower or equal to the end-to-end estimate. In this case, congestion window can be set optimally to match the cellular link bandwidth, thereby ensuring best throughput and minimum delay (due to avoiding queue buildup at the edge). On the other hand, the end-to-end bandwidth estimate lower than the KPI based bandwidth estimate for the cellular link suggests that bottleneck is in the wired segment of the path. In that case, BISCAY is capable of falling back to any wired-specific CCA to deal with the congestion window during the time the bottleneck lies in the wired segment. Note that the overhead associated with simultaneously tracking two bandwidth estimates is negligible, only requiring a few extra bytes of memory and addition/bit-shifting operations. Similarly, switching between two modes (typical and fallback) also has no adverse effect on performance.

Unlike in WiFi, where users compete for the shared resources at the access point (generally in a round-robin fashion), in mobile networks, the base station buffers are not shared among the users, each user having their own isolated deep buffer (no inter-user competition/unfairness). However, those isolated buffers are indeed shared among all the flows from that user device, creating an *inter-flow fairness problem*, given that the different flows of the device will compete to get more resources. This fairness issue is a fundamental problem in the design of all CCAs. Each flow operates a separate independent instance of the CCA, which means that the CCA will determine a congestion window only using that flow's context. The result is an inter-flow competition at the user device, reflecting as unfairness (some flows getting more resources than others) and performance degradation (tail delays). To mitigate this issue, BISCAY takes a global view when determining the congestion window for a given flow on a device. BISCAY keeps track of the number of active flows in the system and uses that number to apportion the calculated bandwidth. Note that the KPI derived bandwidth represents the potential bandwidth that the base station can grant to the UE as a whole, across all its flows. Biscay leaves the strategy used to split the available bandwidth across the active flows up to the implementation – many of the traditional scheduling algorithms proposed in the literature can be leveraged for this purpose.

The foregoing discussion applies perfectly to the Biscay CC operation in the uplink direction. For the downlink direction, where the UE receives data from a remote sender, while some mechanisms underlying Biscay work as is while others need to be adapted, as outlined below. The wireless link bandwidth determination on the cellular downlink using downlink relevant radio KPIs, as described earlier in §3.2.2, can be readily used. On the other hand, the additional capability is required to support downlink CC to relay the cellular link bandwidth estimate to the remote sender. Good news is that the built-in flow control mechanism in TCP offers a readily deployable solution to this issue (this mechanism lies outside the congestion control layer). While the intended purpose of this flow control mechanism is to prevent causing overflow of buffers on the receiver endpoint (UE in this case), the same mechanism can be leveraged to throttle the sender to limit the injected traffic to be limited to the cellular link bandwidth. Note that Biscay's scope is limited to modifications within the UE. In order to used the relayed cellular downlink bandwidth through flow control to also ramp up the sending rate (not only throttling the sender), modification to the sender's TCP stack is needed. In fact, this approach has been employed by prior downlink focused cellular CC proposals [102, 106]. With this feedback mechanism, we can ensure that sender quickly slows down when the cellular link becomes a bottleneck, thus avoiding queue buildup while fully using the link. Ramping up the sending rate in the absence of such a bottleneck, on the other hand, would follow the standard way.

## 4 IMPLEMENTATION

The implementation of Biscay has two core components: the *KPI extraction layer (OpenDiag)* and the *Biscay CC module*. We implemented both these components as kernel modules ($\approx$ 2500 lines of C code). We have developed multiple versions of OpenDiag for different Android versions. At the time of writing, OpenDiag has been tested and validated on Android 11, 12, 13, and 14. But given our extensive experience working with Android 11, our description and experiments are based on that version; it runs on top of the Linux Kernel 4.15 (default for Android 11 on Google Pixel 5). Besides, OpenDiag has also been validated on multiple Android devices and external modems such as Nexus 6P, Samsung Galaxy Note 4, OnePlus Nord 5G, OnePlus Nord N30, Nothing Phone 2, and Quectel modems. OpenDiag has also been used commercially for over a year to collect cellular data across 20 different countries spanning America, Europe and Asia. Additionally, for the sake of simplicity, we package these as part of a Custom Android update, facilitating deployment on non-rooted devices via a manual update. The method for manually updating the OS using a custom image varies across device vendors but in general, this can be done from the Android settings, through an App provided by the vendor or using Android's Fastboot mode.

**KPI extraction layer.** Biscay's KPI extraction layer is named OpenDiag and consists of a multi-threaded kernel module. The first kernel thread is responsible for obtaining and processing data received from the Diag module. This thread handles packet parsing and extracts any KPIs that the CC specifies. The second kernel thread is responsible for the control channel. It is configured to periodically instruct the modem to drain the internal buffer where the diagnosis messages are queued. The control thread flushes the internal buffer every $1ms$ – this frequency was chosen to be smaller than the most frequent diagnostic message, which is generated every $10ms$.

Although Biscay will use OpenDiag from within the kernel, OpenDiag offers a subscription-based API for both kernel and user space consumers. This API is used by the CC layer (the consumer) to specify a list of KPIs that OpenDiag must extract and forward in real-time. The KPI forwarding is done through a shared memory channel in order to avoid excessive message passing within the kernel. This channel is realized through a shared buffer (memory-mapped to the user space application's virtual address space in case of accessing it from user space) that is allocated by OpenDiag based on the number of KPIs specified by the consumer. To avoid unnecessary busy waits (spinlocks), OpenDiag implements triple buffering over the shared-memory region so that the consumer can read KPI asynchronously without causing any race condition. To facilitate user-space interaction (such as user-space TCP or QUIC), we implement *libOD*. libOD exposes a simple POSIX-style API for applications to access radio KPI readings. libOD maintains a set of supported KPIs that can be easily extended to support obtaining any data from any diag message. At the time of writing, libOD supports the majority of relevant KPIs from PHY and MAC layers, the entire RRC and NAS layers (including ASN.1 and L3 Tabular decoding) and a subset of RLC and PDCP layers. Importantly, libOD API has been designed to be compatible with MobileInsight parsers, enabling them to be deployed on top of OpenDiag if needed even though the set of KPIs supported by OpenDiag is larger than what MobileInsight offers. Furthermore, the accuracy of the measurements performed with OpenDiag has been validated against state-of-the-art commercial tools such as Qualcomm's QXDM [82] and Keysight's Nemo Handy [53].

**Biscay CC module.** Biscay has been developed from scratch as an independent CC module only borrowing the pluggable features described in §3 (end-to-end bandwidth estimation and fallback mechanism) from an existing CCA in order to reduce development time. We leverage BBR [34] to implement Biscay's pluggable functionality. In particular, we leverage BBR's *end-to-end* bottleneck bandwidth estimation as part of bottleneck bandwidth estimation as well as vanilla BBR for the fallback mechanism (when the bottleneck switches to the wired segment, vanilla BBR will be used). Our decision is to use BBR's built-in end-to-end bandwidth estimation and fallback is based on the fact that both existing works [106, 108] and our own evaluations show the superior performance of BBR compared to other end-to-end approaches in the wired segment. Other recent works [35, 112] also acknowledge this and leverage BBR as the underlying base framework. These functionalities are taken from the BBR version integrated within the Linux kernel 4.15. A detailed comparison between

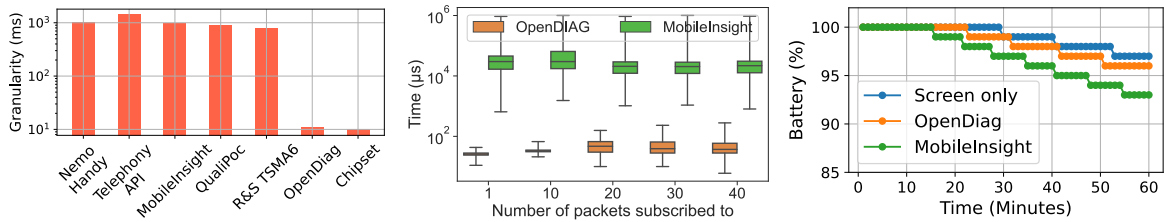

**Figure 4: a)** Granularity of different tools vs chipset (ground truth). **b)** Packet processing time of OpenDiag vs MobileInsight as a function of subscribed packets. **c)** Battery cost over time on COTS UE for Screen, OpenDiag and MobileInsight; both recording the required KPIs.

Biscay's code and BBR's code can be found in §A.5 along with a description of Biscay's internals.

Our implementation complies with the Linux kernel's TCP machinery and gets loaded in the kernel as a CC callback which gets invoked when certain congestion events (e.g. on receiving an ACK, timeouts, duplicate ACKs, or Explicit Congestion Notification) trigger it. Biscay CC module interacts with OpenDiag using the latter's API and converts the KPIs into available bandwidth in Mbit/s. Subsequently, this bandwidth is converted into the bandwidth format (packets/$\mu s$) to further translate it into a congestion window length after multiplying it by the estimated RTT. Biscay lists the active flows, identifying each flow by the 4-tuple corresponding with source IP, destination IP, source port, and destination port. In order to achieve active flow tracking, Biscay overwrites, using a transparent intune, some callbacks of the kernel's socket structure to get notified when any operation is performed in all system's sockets. Given the nature of TCP flows in Android [83] (over 80% of the TCP flows have a lifespan over 10 seconds), we have opted for a bandwidth distribution policy where every flow receive an equal share of the available cellular bandwidth targeting a fair distribution of the available resources. However, we acknowledge and discuss in §A.5 that some scenarios could be negatively affected by this policy and propose alternative scheduling policies that would perform better in such scenarios.

Despite using functionality from BBR, we further discuss how to leverage/integrate other CCAs features within Biscay in §A.5 inspired by how BBR v2/v3 leverages extra signals (ECN and packet loss) from other CCAs [98] to complement its congestion window calculation method.

In addition to this Android implementation, we have also implemented an offline version of Biscay, for evaluations with the Pantheon emulator [110]. Given that Pantheon can only be executed on a computer (we used Ubuntu 18) and OpenDiag cannot run there (there is no modem and Qualcomm drivers), we implemented a trace-based version of Biscay that takes a bandwidth trace generated from KPIs and replays it within Pantheon, effectively mimicking what would happen in the real world within the device.

## 5 EVALUATION

We conduct a comprehensive evaluation of Biscay. First, we evaluate OpenDiag, the KPI extraction layer used in Biscay in isolation, looking at three key parameters: granularity, performance and battery consumption. Then, we evaluate the accuracy of two bandwidth determination methods proposed for Biscay, identifying the optimal granularity. Finally, we compare Biscay with 10 other CCAs under different mobility and workload scenarios. All our experiments are conducted in networks with 4G and 5G coverage.

### 5.1 Biscay's KPI extraction layer

In this section, we discuss the key performance features of OpenDiag, our KPI extraction layer that plays a crucial role in the bandwidth determination accuracy. All the OpenDiag related experiments shown in this section have been conducted over commercial cellular networks.

**Granularity.** Granularity, is the frequency at which packets are received from the chipset, is a key feature that enables real-time data collection. Having a fine granularity means that you can extract samples of a given KPI at any moment in time more accurately. Figure 4a shows a comparison between six different collection tools (Nemo Handy [53], Telephony API [45], MobileInsight [66], QualiPoc [84], R&S TSMA6 [85] and OpenDiag) and the finest granularity at which the chipset can report a given KPI which is the ground truth. For this experiment, we chose Reference Signal Received Power (RSRP) as the extracted KPI. RSRP is a standard KPI used in multiple works, and it is one of the KPIs present in the most frequent packet generated by the chipset through the diag interface (LTE Serving Cell Measurement Response packet).

In Figure 4a, granularity is reported as the time between samples in $ms$. While the minimum packet granularity offered by commercial [53, 84, 85], standard [45] and open source solutions [66] is in the order of 1000$ms$, OpenDiag is able to retrieve packets from the chipset almost every 10.9$ms$, a 100× improvement over all the alternative tools. Despite it's in-kernel design, the main reason behind such an improvement is the use of the control channel of OpenDiag that forces the chipset to drain the packets of the modem's internal buffer every $ms$. A similar improvement (95×) has been obtained using a user-space version of OpenDiag, which rules out the kernel factor as the main reason for the improvement. The overhead created by the message gathering and parsing processes generates a 9% overhead over the ground truth with 10$ms$ granularity. Interestingly, if the logs generated by MobileInsight are analyzed, the reported granularity matches the ground truth. However, this alignment is deceiving due to MobileInsight's time mechanism which uses the timestamp that comes in the header of each packet (the time at which the chipset created that packet) as the packet timestamp rather than the time at which the packet is received by the application. In practice, if MobileInsight is used and the time at which the application receives a packet is recorded, we will observe that a batch of packets is received at a given time due to the chipset's internal buffering. Roughly one second later, another batch of packets is received by the application, comprising those generated during that particular second. Extending this behavior to the value of a given KPI, we will observe that the value of the KPI remains unchanged for a second and, when the batch of packets is received and parsed, its value will change several times within a period of a few nanoseconds. In practice, this is seen as one-second granularity.

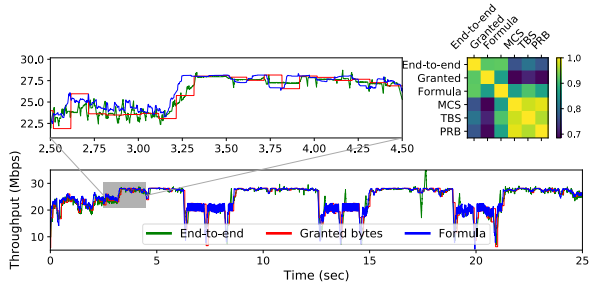

**Figure 5:** Throughput comparison of granted KPIs and 3GPP formula with end-to-end ground truth. Heatmap indicates the high Pearson correlation in the bandwidth determination methods.

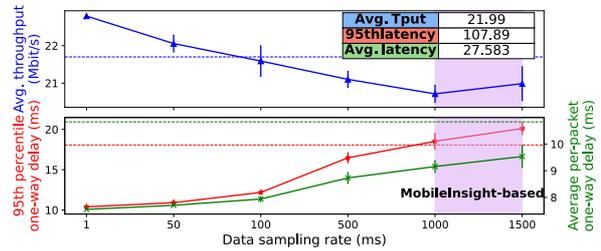

**Figure 6:** Effects of data sampling frequency on different transport metrics. The table on top shows BISCAY's performance using the average channel bandwidth.

**Processing time.** Besides the granularity, the other key factor that prevents us from using MobileInsight as the KPI extraction layer is the processing performance rooted in its multi-layer design and inefficient processing pipeline.

In Figure 4b, we measure how long each packet spends in the processing pipeline of OPENDIAG and MobileInsight. This is the time between the packet's arrival in the processing pipeline and the time it has been dispatched and is available for the consumer. Even though BISCAY only uses a few KPIs, we conducted a comprehensive evaluation where we extract one KPI for every packet that we are subscribed to while we increase the number of packets. Please note that each packet type might contain multiple KPIs. Due to the limitations of MobileInsight's data forwarding to a third-party application, we implemented a dummy consumer within itself (OPENDIAG's consumer was another application, incurring extra time to forward the KPIs). From the numbers shown in Figure 4b, it is clear that OPENDIAG's design plays a significant role in terms of performance, i.e., timely KPI retrieval with improvements in the order of $100 - 1000 \times$. While OPENDIAG's additional delay remains in the order of tens of $\mu s$, MobileInsight's pipeline generates processing delays of hundreds of ms, making it completely unusable to calculate the grants received in real-time every TTI (5G TTI is $0.5ms$).

**Energy efficiency.** BISCAY requires a KPI extraction layer to run in the background in order to calculate the bandwidth. Due to this, we conduct an evaluation of the battery life penalty that the user must pay for using OPENDIAG as a KPI extraction tool compared to MobileInsight. We discover that the modem only generates debug packets when it is in active mode. During the low-power mode, it generates very infrequent and periodic reports. So, to keep the UE in active mode, we are generating a small but constant data plane traffic using `iperf3`. Besides, for all our tests, we disabled Android's adaptive battery and screen brightness options to maintain consistent behavior between measurements.

Figure 4c shows the battery consumption over an hour of three different configurations: Screen only (the screen remains on while `ping` was running in the background), OPENDIAG (the screen was on and the required KPIs for BISCAY were recorded using OPENDIAG), and MobileInsight (the required KPIs were recorded using MobileInsigth). For all three configurations, the modem was in active mode. The results show that the battery degradation of OPENDIAG compared with the baseline (Screen only) is negligible over the 60-minute period. However, with MobileInsight, the battery consumption is more noticeable. We mainly attribute this to the multi-layer design used by MobileInsight, which requires multiple applications to run concurrently.

### 5.2 BISCAY's bandwidth determination

The design section (§3) introduces two different ways of determining the maximum available bandwidth in the radio link: the simplified 3GPP throughput calculation and MAC layer granted bytes. Theoretically, the main advantage of the former method comes with its granularity (throughput can be determined at TTI granularity) with the trade-off of extracting multiple KPIs such as PRBs, TBS index, and MIMO for all the serving cells if CA is enabled. On the other hand, the advantage of using granted bytes comes from its simplicity (one single KPI contains the resulting throughput after considering 4G/5G, CA, MIMO, etc.); however, this KPI gets updated every $100ms$.

Figure 5 shows a correlation study between the two throughput determination methods defined by BISCAY with the ground-truth throughput that corresponds with the throughput received at the receiver side (server). We also include the raw KPIs used in similar works [63, 104, 106] to estimate the bottleneck bandwidth (BISCAY uses those KPIs as indexes in the pre-calculated 3GPP-defined tables). The correlation matrix clearly shows an extremely high Pearson correlation (over 0.95) with the bottleneck bandwidth of the two methods defined by BISCAY. Interestingly, even though it suffers from a $100ms$ granularity, the granted bytes KPI can perform as well as the simplified 3GPP formula, suggesting that having extremely fine granularity (TTI-granularity) might not be a decisive factor in CC. This experiment also proves how simply using raw KPIs is not an accurate method to determine the radio link throughput, with the correlation coefficient barely reaching 0.75. A time-series plot of a randomly picked scenario (mobility on-peak with CA enabled and 4G+5G) complements the correlation matrix showing how similar to the ground truth both throughput determination methods

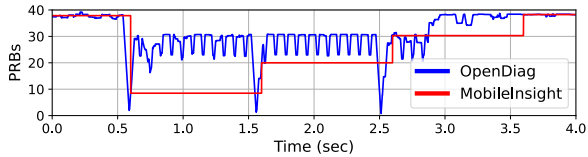

**Figure 7:** Measurement of PRBs with OpenDiag & MobileInsight.

are. Please note the propagation delay has been removed for visualization purposes.

Figure 7 illustrates the time series of PRB allocations obtained using both OpenDiag and MobileInsight. Note that PRB is a common KPI used in similar works [63, 104, 106]. Due to the differences in update frequencies between these approaches, we observe a clear distinction in the measured KPI (PRB) time series. The less frequent updates with MobileInsight causes applications relying on its measurements (e.g., PERCEIVE [63] or Claw [104]) to rely on outdated or less precise estimates during intervals between the updates. In contrast, the fine granularity provided by OpenDiag leads to accurate measurements, enabling more effective and optimal utilization of resources. The discrepancy between reported and actual real-time KPI data can significantly impact the performance and efficacy of applications reliant on KPI based estimates.

The combination of the proposed bandwidth determination method's accuracy (Figure 5) and precise KPI measurements resulting from a finer granularity sampling (Figure 7) hints towards optimal performance from transport layer point of view. Given this, we decided to conduct an experiment to identify the effects of the data granularity on the transport layer performance metrics (throughput, average and tail delay at of 95th percentile delay). The results depicted in Figure 6, show the variation of end-to-end transport layer metrics with the KPI sampling interval. We modified BISCAY to sample the air interface at a given pre-defined frequency varied between $1 - 1500ms$. The scenario shown in Figure 6 is the same scenario shown in Figures 5 and 7. We added the results obtained by BBR as dashed horizontal lines as reference.

The results clearly show that while the throughput does not decrease significantly as we increase the sampling interval, both average and tail delays do increase with the KPI sampling interval. This behavior implies that coarser sampling rates (KPI granularity) translate to BISCAY saturating the channel, leading to queue buildups resulting in the same throughput and more delays. Interestingly, this behavior is only noticeable for sampling rates larger than $100ms$, which explains why there is no difference between the throughput determination methods evaluated in the correlation analysis. For sampling rates higher than $100ms$, BISCAY starts to perform like BBR and even underperforms BBR in both throughput and delay if the granularity is coarse enough. Both, throughput and delay performance get degraded even more when granularity beyond $1000ms$ is used (where all MobileInsight-based solutions [63, 104] operate). We have added a table in Figure 6 (at top) that shows BISCAY's performance using the coarser sampling rate, sampling the channel just once.

If this experiment is repeated multiple times, the result will be the same as using the average channel throughput in BISCAY. Consistent with the graph, the throughput remains within a constant range while the tail delay and average delay grow drastically to $107ms$ and $27ms$, respectively. These results and the required granularities highlight the need for a KPI extraction layer like OpenDiag given that the best granularity that can be obtained with all the alternative solutions is $1000ms$. It also validates that there is no significant difference in transport performance between the two throughput determination methods besides the implementation complexity. Therefore, our presented implementation uses the MAC layer granted bytes KPI as a proxy to calculate the radio link bandwidth.

### 5.3 Congestion Control

In this section, we explore the performance of BISCAY in terms of throughput and delay (average and tail delays) in commercial 4G and 5G networks. We compare BISCAY with 8 other state-of-the-art CCAs. We deliberately picked a representative of each of the categories discussed in the related work for the comparison: CUBIC [49] (loss-based), Copa [29] & LEDBAT [87] (delay-based), BBR [34] (hybrid), PCC [36] & Vivace [37] (learning-based), and Sprout [102] & Verus [111] (wireless-aware). As an additional baseline, we include the Oracle, a CCA that precisely knows the end-to-end bottleneck bandwidth at any given point in time and so achieves the optimal performance (maximum throughput and minimum delay).

*5.3.1 Real World Dataset Collection.* To record the traces, we measured the uplink and downlink throughput across time by saturating the link (both directions individually) with MTU-sized UDP packets. This is because TCP cannot reliably saturate the channel as CC will kick in and reduce the sending rate. This methodology is consistent with prior research [86, 99, 102]. We recorded the throughput within the UE (Google Pixel 5, OnePlus Nord N30 and OnePlus Nord 10T) and in the receiver server using tcpdump to ensure the correctness of our measurements. iperf3 was used within sender as a traffic generator. To extract the required KPIs, OpenDiag was running in parallel with tcpdump in the UE. Those KPIs are later introduced in a trace that is fed into Pantheon §5.3.2. This setup was used to collect multiple traces under four different scenarios: mobility on-peak, mobility off-peak, static on-peak and static off-peak. The on-peak label corresponds to a trace collected during busy hours (9 am to 6 pm), whereas the off-peak label corresponds to traces collected between 10 pm and 2 am. We define a scenario as mobile when the UE is moving between cells (labeled as mobility traces recorded while walking, driving, bus, and train), and static corresponds to when the UE is not moving between cells. All data collection occurred within an urban or campus area in major cities in EU and US.

For a robust evaluation, we intentionally designed our data collection approach to transition between 4G and 5G (NSA and SA) covered areas during each measurement. So, every mobility trace includes both 4G and 5G data. Additionally, the mobility traces were recorded with CA enabled, reflecting the effect of being served by multiple cells. We collected multiple traces in different locations for the static case to reflect the diversity in network conditions (4G/5G and single/multiple serving cells). Our methodology aimed to provide a well-rounded view, ensuring the reliability and generality of our results.

More information about the dataset, its characteristic and scenarios distribution can be found in A.4

### 5.3.2 Testbed Configuration.
The evaluation is primarily conducted on the Pantheon emulator [110], a network emulation tool replaying pre-recorded network traces under emulated network conditions. Pantheon is built on top of mahimahi [73], another emulation framework initially designed for HTTP-based traffic. Both of these tools are widely used in the networking research community [29, 102, 106, 109, 110]. Pantheon is deployed in the Powder platform [6, 32] using a 32-core CPU machine with 64 GB of memory running Ubuntu 18. We generated 260 distinct Pantheon traces from our measurement campaign. The reason behind using Pantheon for a subset of the experiments is to ensure a fair evaluation where all CCAs are evaluated under the exact same conditions since that cannot be guaranteed in the wild.

### 5.3.3 Single-flow performance.
Figure 8 compares Biscay with the eight CCAs mentioned above under four scenarios: Mobility On-Peak, Mobility Off-Peak, Static On-Peak and Static Off-Peak. The x-axis of the plot is reversed and the top right region is the best performing throughput-delay pair. Each graph is the average from the respective set of traces. It is important to highlight that Pantheon will determine the emulator delay based on the provided trace, which is not representative of the delay experienced during the measurement campaign. However, the queue delay ratio amongst the different CCAs is trustworthy.

Across all the scenarios, CUBIC has the highest delays (average and tail) because its cwnd gets reduced only when a loss is detected. Until then, the bottleneck queue builds up, leading to increased delays before the packets are dropped. Copa and LEDBAT, both delay-based, report throughputs similar to CUBIC and a fraction of CUBIC's delays across all the scenarios.

Sprout and Verus can significantly reduce the delay, given that they were specifically designed for wireless access networks. Interestingly, Sprout matches CUBIC's throughput, but Verus is far from that, suggesting that the source of its low delay (although extremely high tail delay) is due to not fully saturating the channel. Learning-based CCAs (PCC and Vivace) show abnormal and inconsistent behavior across the different scenarios for both throughput and delay, hinting that the models employed are overfitting in some scenarios. BBR results outperform all the previously discussed CCAs in all the scenarios, both in terms of delay (excluding overfits and abnormal results) and throughput. Finally, Biscay is able to maximize the channel usage, resulting in maximum throughput without paying delay penalties attributed to its accurate bottleneck bandwidth determination method that calculates the precise bandwidth in real-time and adjusts the sending rate accordingly. Moreover, among all the CCAs evaluated, Biscay is the closest to the Oracle, both in terms of throughput and delay in all the scenarios. The reason for this lies in the fact that the different bandwidth calculations used by the evaluated CCAs are inaccurate and coarse-grain approximations of the bottleneck bandwidth. Finally, an interesting observation is that, on average, all the evaluated CCAs seem to have better performance (more throughput and equal or lower latency) in mobile scenarios as opposed to static, which is counterintuitive. That is just an artifact of the dataset used; more on this in A.4

| CCA | Scenario | Two flows | | | Three flows | | |
|---|---|---|---|---|---|---|---|
| | | Avg Tput (Mbit/s) | Avg Delay (ms) | Tail Delay (ms) | Avg Tput (Mbit/s) | Avg Delay (ms) | Tail Delay (ms) |
| Biscay | Mobile | 12.76 | 5.625 | 7.64 | 8.5 | 5.78 | 7.54 |
| | Static | 8.89 | 10.92 | 13.62 | 5.91 | 10.04 | 12.79 |
| BBR | Mobile | 12.59 | 7.68 | 9.73 | 8.42 | 7.91 | 10.05 |
| | Static | 8.46 | 14.35 | 19.69 | 5.54 | 17.01 | 25.15 |
| CUBIC | Mobile | 13.54 | 417.2 | 802.9 | 9.17 | 397.33 | 830.38 |
| | Static | 9.13 | 768.9 | 1536 | 5.91 | 667.34 | 1450.80 |

**Table 2:** Performance comparison with two and three simultaneous flows using Biscay, BBR and CUBIC.

### 5.3.4 Multi-flow performance.
For a comprehensive analysis of Biscay's performance, we conduct an experiment where multiple flows are simultaneously active. From the experiments shown in Figure 8 and summarized in Table 1, it can be seen that BBR is closest in performance to Biscay across all the scenarios. Additionally, CUBIC is the default CCA in the Linux operating system and, therefore, in Android. We consider these two CCAs for the multi-flow evaluation. The performance of the CCAs was evaluated under multiflow conditions by simultaneously running two and three flows (Table 2). The idea behind this evaluation is that an end device hardly runs a single flow and is configured with a given CCA, which gets applied at the OS level. Therefore all the TCP sockets open in that system will use the predefined CCA unless otherwise specified through the socket options, a practice rarely seen outside networking laboratories. For simplicity's sake, both tables contain the average results of On-Peak and Off-Peak for the Mobile and Static scenarios.

We observe similar behavior to that of single-flow experiments in terms of throughput. All three CCAs saturate the channel, and there is not much throughput difference between them. However, not all of them maximize channel usage at the same cost. CUBIC saturates the channel, which is reflected in high average and tail delays (up to 1.5s). BBR's more precise bottleneck bandwidth estimation leads to smaller queues at the bottleneck link, resulting in lower average and tail delays. However, BBR design does not consider other in-device flows when determining the cwnd of a given flow, which is reflected when flows compete for bandwidth, resulting in queue buildups and higher delays (average and tail) increases. BBR is more conservative and decreases the window whenever it detects queues being built. On the other hand, Biscay is aware of the number of active flows in the system, and it is able to divide the `grant received` by the base station, which is meant for the entire device evenly among the active flows avoiding the competition and limiting the queue buildup. Besides the apparent decrease in throughput, the flow-aware design of Biscay enables it to maintain similar or even lower average and tail delays. This demonstrates the efficacy of Biscay's design in managing multiple flows.

### 5.3.5 Fairness.
Building on the multi-flow experiments where flows competed for bandwidth, we evaluated how fair Biscay is under such competition and compared it with BBR and CUBIC. Unlike WiFi networks, where devices contend for access to the same physical resources, in mobile networks, the base station manages the allocation of resources to the users. The base station's MAC scheduler makes resource allocation decisions based on the Scheduling Request received from the UE and the observed channel quality. The scheduler then grants a portion of the available resources to the UE through a Downlink Control Information (DCI) message. In commercial RANs, scheduling algorithms typically use a proportional fair approach to ensure fairness among users, as fairness is a crucial requirement in the RAN [57]. Additionally, unlike WiFi access points, where all users' data is queued in the same buffer, base stations have dedicated deep buffers for each UE. Therefore, inter-user fairness relies on the base station scheduler, not the UEs. However, an inter-flow competition where multiple flows compete for the uplink bandwidth within the UE is still an issue that CCAs have to deal with.

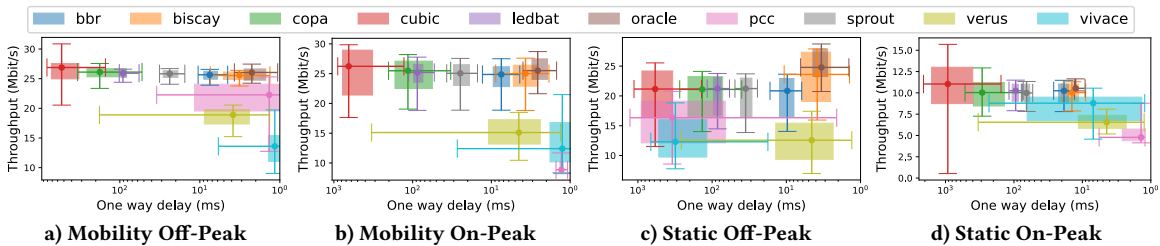

**Figure 8:** Single-flow throughput and one-way delay (axis reversed) comparison of different CCAs in commercial networks. The boxes' horizontal edges (one-way delay) represent the 25th and 75th percentiles, and the ends of the error bar give the 10th and 90th (tail delay). Similarly, vertical box edges represent the 25th and 75th throughput percentiles, and the error bars give the 10th and 90th percentiles. The error bar intersection shows the throughput and delay medians. The top-right is the best performing throughput-delay pair.

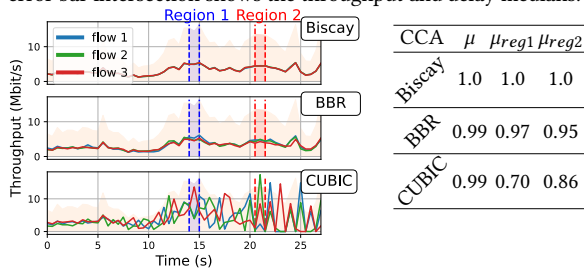

**Figure 9:** Per-flow throughput for BISCAY, BBR and CUBIC. The shaded region indicates the total channel bandwidth. **Table 3:** Jain's fairness index of the different shadowed areas.

| CCA | $\mu$ | $\mu_{reg1}$ | $\mu_{reg2}$ |
|---|---|---|---|
| BISCAY | 1.0 | 1.0 | 1.0 |
| BBR | 0.99 | 0.97 | 0.95 |
| CUBIC | 0.99 | 0.70 | 0.86 |

Figure 9 shows the throughput achieved by three simultaneous flows when running BISCAY, BBR and CUBIC as time series. The shaded area in the background represents the total channel bandwidth. In CUBIC, each flow strives to maximize its throughput, leading to a suboptimal allocation of resources. In contrast, BBR is more conservative, focusing on reducing delay, resulting in a more uniform distribution of the channel resources among active flows compared to CUBIC. Finally, BISCAY is able to split the available bandwidth equally among the existing flows, which is attributed to having a global view of the bandwidth available for the UE, ensuring that every flow gets the same amount of bandwidth from the total available. Additionally, we have calculated the fairness using Jain's fairness index [51]. Figure 9 (right) includes the fairness index of the entire experiment and the two shaded areas (Region 1 and Region 2). One of the limitations, as can be seen here, is that Jain's index uses the average throughput over the selected time period masking and hiding fine-grain details. This is proven by the fact that the fairness index of the entire experiment is near perfect for the three CCAs with an index equal to 1, while the time series show contradictory behavior. The fairness indexes of the two shaded regions show that even though the three CCA might look fair over long periods, BBR and CUBIC bandwidth distribution is unfair (particularly harmful in short-duration communications such as web traffic).

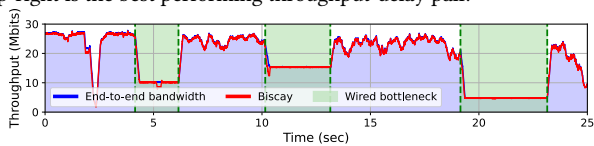

**Figure 10:** BISCAY falling back to BBR when the bottleneck changes to the wired segment

*5.3.6 Bottleneck detection.* Finally, we assess BISCAY's ability to detect changes in bottleneck location and its intended behavior of falling back to a wired-specific CCA (our implementation uses BBR) when the bottleneck shifts to the wired segment of the path. Although we did not encounter any instances of wired-segment bottlenecks in our measurements and experiments on public networks, we intentionally simulated this scenario. This involved manually limiting the bandwidth of the wired segment to a value lower than that of the wireless link. As shown in Figure 10, we realize this scenario by setting the end-to-end bandwidth to 5, 10, and 15 Mbps at arbitrary times throughout the experiment. Leveraging the mechanism described in §3, BISCAY detects when the end-to-end bandwidth reduces compared to the KPI-based wireless link bandwidth estimate and swiftly switches to BBR. Conversely, when that condition is not met, BISCAY goes back to its normal operation mode. This prompt adaptation to changes in the network conditions ensures an effective utilization of the available bandwidth.

## 5.4 Wireless-aware CC Evaluation

So far, we have compared BISCAY with the most used and relevant CCAs; however, there are few wireless-specific CCAs that also leverage air-interface KPIs to operate that are missing in the evaluation. Specifically, PBE-CC [106], a CCA built atop NG-Scope [105] which is LTE-only sniffer that extracts DCI messages from the air interface containing the scheduling grants of the users in a given cell. PBE-CC works on the principle of exploiting all the *available* PRBs that have not been allocated to any user in the air interface to transmit data. Note that PBE-CC operates only in 4G downlink direction, thus, we have implemented an equivalent version of BISCAY for a fair comparison. Inspired by the feedback mechanism employed in prior wireless/cellular CCAs such as Sprout [102], in our implementation, the cellular downlink bandwidth derived from downlink KPIs (PRBs and TBS extracted from the DCI downlink grant) is sent to the sender using the TCP flow control mechanism. BISCAY calculates the observed downlink bandwidth using OPENDIAG and sends it to the other endpoint which uses it to determine how much traffic is sent.

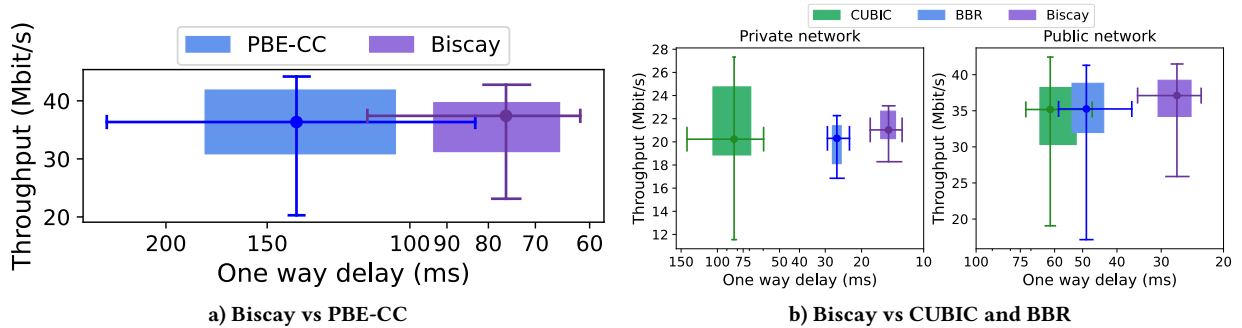

**Figure 11: a)** Throughput and delay comparison of BISCAY compared with PBE-CC. **b)** Throughput and delay comparison of BISCAY, BBR and CUBIC using a COTS UE (Google Pixel 5).

Figure 11a demonstrates the comparison of BISCAY and PBE-CC in commercial 4G network scenarios. We observe that both PBE-CC and BISCAY achieve similar throughput as both maximize the available network bandwidth. However, BISCAY outperforms PBE-CC in terms of delay with BISCAY halving the average and tail delays compared to PBE-CC. This behavior can be attributed to PBE-CC's mechanism of increasing the downstream sending rate when PRBs are available and assuming that the base station will grant those resources to it. This simplistic and naive view of the base station's MAC scheduler is far from how commercial schedulers work. In practice, however, most if not all schedulers implement some variation of proportional fair (PF) which uses the number of UEs in the cell and the UE scheduling requests/buffer status reports (a reflection of how much data each UE wants to transmit), channel quality measurements (CQI and periodic measurement reports concluded by the UE and the base station), bearer quality of service and even the historic grant allocation as input. The performance of PBE-CC is intricately linked to the channel quality reports from both the User Equipment (UE) and the base station. The base station grants resources based on these quality assessments: a poor channel quality results in fewer PRBs with lower MCS to ensure the UE can decode them. While UEs typically receive resources if they have data to send, boast good channel quality, and haven't recently received resources. PBE-CC's narrow focus on only the available resources ignoring everything else can lead to queue buildups, thus increasing delays.

### 5.5 BISCAY's Real-World Evaluation

Pantheon emulation can deviate from the real-world throughput and delays by up to 17% [110]. To demonstrate its deployability and to overcome the inherent limitations of emulation, we implemented BISCAY on a COTS UE (Google Pixel 5). We evaluated it alongside BBR and CUBIC on real private and public networks. Real-world evaluations pose a significant challenge due to variable external conditions, like random noise and other users in the cell across experiments. Such factors can significantly impact the performance of the different

CCAs. To minimize such external factors across experiments, we used a private network with two base stations (leveraging srsRAN [93]) configured to be neighbor cells (with handovers between them) and core network (Open5GS [64]). To ensure reproducible network conditions, the experiments were conducted when there was no other COTS UE in the network. In addition to private network deployment, we also conducted evaluations on public networks of two major US operators (Verizon and T-Mobile). These evaluations span diverse scenarios, including: static/mobile, single-cell/CA, 4G/5G networks, and on-peak/off-peak periods. For consistency, the same mobility patterns and static positions were used across experiments in both private and public networks.

Figure 11b shows the results of experiments using a COTS UE with CUBIC, BBR and BISCAY over both private and public networks. Similar to the results obtained with Pantheon, CUBIC saturates the bottleneck wireless link, resulting in maximum bandwidth usage and high average/tail delays due to queue build-up. Compared to CUBIC, BBR achieves similar average throughput, but has lower average and tail delays due to relatively more accurate bottleneck bandwidth estimation. Notably, BISCAY outperforms both BBR and CUBIC in terms of throughput and delay in both private and public networks. Specifically, BISCAY gets 4.6% higher throughput than CUBIC and BBR while reducing average and tail delays by 46% and 44%, respectively, compared to BBR. Interestingly, the highest delays experienced by BISCAY are lower than the lowest delays experienced by CUBIC and BBR. This experiment demonstrates how the different components of the BISCAY system effectively work together in challenging real-world conditions.

## 6 CONCLUSIONS

We propose BISCAY, a practical and radio KPI-driven congestion control design for mobile networks. BISCAY leverages OPENDIAG, our in-kernel real-time radio KPI extraction tool that allows KPIs to be obtained from the radio modem at fine *ms* scale granularity. It enables BISCAY to accurately determine the bottleneck bandwidth on the device side to achieve high throughput and low delays. BISCAY is extensively evaluated and compared against 9 state-of-the-art CCAs in a wide variety of scenarios using our 4G/5G performance traces and real-world experiments using a commodity mobile device. BISCAY shows a significant reduction in average and tail delays, notably 58%/41% and 98%/99% average/tail delay reduction compared with BBR and CUBIC, respectively.

**Ethics.** This work does not raise any ethical issues.

# A APPENDIX

## A.1 Mobile Network Stack

Both 4G and 5G stacks are quite similar and reside under the IP layer in the TCP/IP model and provide similar functionality. For the sake of concreteness, however, we will focus on the 5G mobile network stack (illustrated in Figure 12).

Starting from the bottom, the Physical layer (PHY) [14] provides a transport channel to the upper layers and transfers higher layer information over the air interface to the 5G base station (gNB). Immediately above, the Medium Access Control layer (MAC) [11] serves as an interface between logical channels and the transport channel at PHY providing data transfer and radio resource allocation services to upper layers. The Radio Link Control layer (RLC) [17] sits on top of the MAC and is responsible for the transfer of upper layer Protocol Data Units (PDUs), error correction, concatenation, segmentation, reordering, duplicate detection and reassembly. Packet Data Convergence Protocol layer (PDCP) [13] – the layer on top of RLC – is responsible for transferring user and control plane data, header compression, and ciphering/integrity protection.

In between the PDCP and the IP layers, the Service Data Adaptation Protocol (SDAP) layer [19], a new addition relative to 4G, is in charge of the user plane traffic's quality of service. On the other hand, for the control plane, the Radio Resource Control (RRC) [18] configures the user and control planes according to the network state and is in control of information of the connection establishment/release, system information broadcast, radio bearer establishment/reconfiguration/release, mobility procedures (handovers) and paging notification. Finally, over the RRC layer, the Non-Access-Stratum layer (NAS) [12] is in charge of the session management procedures (authentication, security control, mobility, etc.) to establish and maintain IP connectivity between the device (UE) and AMF in the mobile core. The data communication between the device and remote endpoint happens via a tunnel to the UPF in the 5G core.

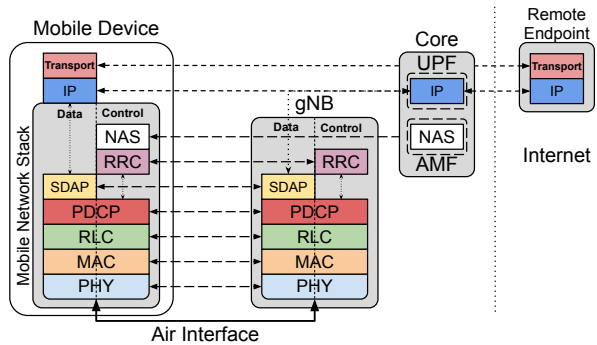

Figure 12: Schematic of 5G mobile network stack on device.

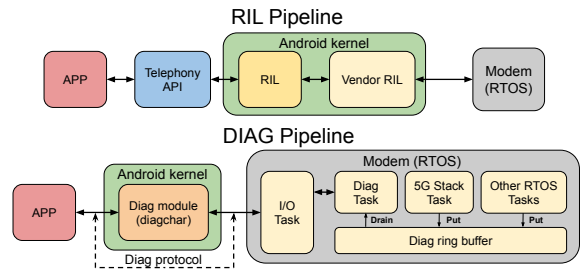

Figure 13: Architectures of RIL [9] (top) and DIAG [40] (bottom).

### A.2 Communication with the Radio Modem

As pointed out in [66], the radio modem in a mobile device comes with a debug/diagnosis channel [10, 27, 38, 71, 81] that is primarily meant for Original Equipment Manufacturers (OEMs) to perform advanced baseband configurations and diagnostics. Here we focus on Qualcomm (Snapdragon series) modems given that they are the most common radio chipset in 5G devices. The diagnostic channel architecture in Qualcomm modems can be generalized to other manufacturers to a large extent.

Every modern Qualcomm system on a chip (SOC) contains two different processing units: CPU and DSP. The CPU (typically an ARM based architecture) runs a general-purpose OS (GPOS) such as Android or iOS, whereas the DSP or modem (usually Hexagon based architecture [7]) runs a real-time OS (RTOS) such as Qualcomm QuRT RTOS. The GPOS and the RTOS are completely isolated from each other, and they can only interact with each other through a standard communication channel. This means that any process running in the GPOS (or even the GPOS kernel itself) cannot access anything within the RTOS or vice versa unless the standard communication channel is used. In Android, this channel is called Radio Interface Layer (RIL) [9], whose use is transparent to the user. The RIL defines a generic interface that applications (and even Android itself) use to interact with the modem. Some examples of the functions provided by the RIL are starting a call, terminating a call, introducing the SIM card pin and getting the coverage level. Given that RIL is a generic and modem agnostic interface, each modem manufacturer must provide a translation layer between the RIL and the specific modem that is referred to as the vendor-RIL, as illustrated in Figure 13. Qualcomm's vendor-RIL uses the standard RIL interface on one side and on the other side QMI (Qualcomm MSM Interface) [8] – a proprietary protocol used to interact with Qualcomm modems. From the perspective of radio KPI data collection, RIL offers only a small subset and that too a coarse time granularity (2-3s) via the Telephony API (a set of libraries built on top of RIL) [45].

Qualcomm modems additionally provide a side-channel called Diag (diag is also the name of the protocol) [77] for diagnostics and control. Unlike the RIL, diag was designed to provide all sorts of debug information and control capabilities so that manufacturers can use it to diagnose the modem using dedicated tools such as QXDM [82]. The only way of accessing the diag functionality is through the Diag kernel module (*diagchar*), an open-source kernel module provided by Qualcomm that acts as a shim between an application and the chipset, and exposes only the basic functionality (read, write and minimum protocol configuration). In practice, the Diag module is just a proxy that simplifies the access to the chipset. The bottom part of Figure 13 shows a schematic of the DIAG architecture. The application must implement the undocumented and proprietary diag protocol logic to communicate effectively with the chipset. Broadly speaking, the diag protocol offers two sets of features: gathering features (read and parse debug messages coming from the chip) and control features (modifying the chip's behavior and state). The former set of features have been partially reverse engineered and implemented by some KPI collection/measurement tools. The latter set of features have not previously been exploited by any measurement tool but enable an application to modify the behavior and state of the chip (change internal variables or disable the internal message buffering).

### A.3 Carrier aggregation

Carrier aggregation (CA) [20] is a technique introduced in LTE-Advanced and remains an integral part of 5G for increasing the per-user bandwidth and the user throughput via aggregation of radio resources in the form of frequency blocks (called component carriers) from multiple cells and assigning them to the UE. CA is used when the amount of data to be transferred for the UE is insufficient with the resources from one cell, Primary Component Carrier (PCC), which is when the base station activates new cells or Secondary Component Carriers (SCC) to cope with that additional load. Though the SCCs are added and removed as needed, the PCC only changes at handover; the UE relies on the PCC for the RRC connection and to send/receive NAS information (e.g., security parameters). CA scenarios are common in dense urban environments where the number of available cells is higher.

### A.4 Dataset and evaluation scenarios

The result of our measurement campaign is a dataset that contains 4G and 5G traces in equal proportion collected in a variety of cities across Europe and the US. Within the 4G portion of the dataset, 50% of the traces are collected in static scenarios and 50% under mobility. However, the 5G part of the dataset is split in a 1 to 4 proportion (20% static and 80% mobility). This distribution may lead to artifacts in the evaluation, such as CCAs appearing to perform better under mobility scenarios. This outcome might seem counterintuitive given that 5G generally offers superior performance compared to 4G, with higher throughput and lower latency. Across both 4G and 5G traces, 60% were collected during peak hours, while the remaining 40% were gathered during off-peak hours.

In addition to those four scenarios, the collected traces as well as the real-world experiments capture the complete range in terms of connection states and events [18]: 4G/5G RRC idle to connected (the modem is disconnected from the network and new connection is established in order to send traffic), 5G RRC inactive to connected (the modem is not fully disconnected from the network and the previous RRC session gets reused to establish a data channel), network attachment and sessions establishments over all the radio access technologies (4G/5G NSA/5G SA) and all possible handover combinations (4G-to-4G, 4G-to-5G, 5G-to-4G and 5G-to-5G). All the previously listed events and state transitions are evenly distributed across the different scenarios captured in the measurement campaign as well as the live experiments.

It is important to note that the aforementioned events and state transitions are orthogonal to the CCA and equally affect all the evaluated CCAs since the radio layer sits underneath the transport layer where congestion control is used.

## A.5 Discussion on BISCAY CC internals

*A.5.1  BISCAY and BBR.* The following text presents the pseudocode for BISCAY's functionality. Like other congestion control algorithms in Linux, BISCAY's CC module gets deployed in the kernel as a module and the congestion control function (*calculate_congestion_window_callback()*) is invoked as a callback. BISCAY logical structure is a state machine with three states: STARTUP, BISCAY, and FALLBACK.

**STARTUP**: When a TCP socket is open, it goes into the STARTUP state/phase. Like other CCAs [49], we leverage slow-start [50] which exponentially increases the congestion window as congestion is not reached in order to fill the pipe quickly. However, unlike other CCAs that remain in the slow-start phase until a congestion condition is met (e.g., the bandwidth reaches a ceiling, packets start dropping, delay increases, etc.), BISCAY exits the STARTUP state and moves into BISCAY when a reliable KPI-based bandwidth prediction can be made (usually after a couple of slow-start iterations). Note that in practice, when the phone has been running for some time, this phase is mostly skipped given that it is highly probable that when a socket gets open and starts transmitting (STARTUP state), other sockets are already transmitting in the system and therefore reliable KPI-based bandwidth predictions from OpenDiag can be obtained.

**BISCAY**: During the BISCAY state, BISCAY sets the congestion window according to the logic defined in §3. The cellular bandwidth (*cellular_bw*) is calculated from the KPIs extracted with OpenDiag and combined after splitting it according to the number of active flows (*BWSplitPolicy(bw)*) with the RTT (obtained as the minimum of the previous RTT value and the RTT of the last ACK) into a congestion window value using the bandwidth-delay product. Then, BISCAY checks for a change in the bottleneck location. It does so by comparing the end-to-end bandwidth (obtained using BBR's end-to-end bandwidth estimation) with the cellular bandwidth. In case the bottleneck moves to the wired segment, BISCAY switches state to FALLBACK and falls back to an end-to-end CCA (BBR). Otherwise, it returns the congestion window previously calculated.

**FALLBACK**: In the FALLBACK state, BISCAY calculates the congestion window using the selected wired-specific CCA (BBR). Then, like in the BISCAY state, BISCAY checks for changes in the location of the bottleneck using the cellular bandwidth and the end-to-end bandwidth and if the bottleneck switches, the congestion window gets calculated using the KPI-derived bandwidth and the state changes to BISCAY.

In the pseudocode, the lines highlighted in red indicate areas where external CCA logic is integrated. For ease of implementation, we leverage BBR.

*A.5.2  Multi-flow bandwidth distribution.* Given the flow's homogeneity [83] (where the average lifespan of flows tends to be similar), we have implemented a simple yet effective bandwidth distribution policy that equally splits the available bandwidth among the active flows. This approach not only promotes fairness but also ensures efficiency by avoiding complex kernel-level computations, particularly floating-point operations, which are unsupported and could introduce additional delays.

Although BISCAY's focus is on TCP, our bandwidth distribution policy also aims for fairness with other transport protocols by accounting for UDP connections (QUIC [61], which is used by a number of Google applications and web-based applications, runs over UDP). It does so by also including UDP active flows when calculating the number of active flows in the system (*getNumberActiveFlows()*), ensuring fairness across transport protocols.

However, there are still some scenarios that are negatively affected by our bandwidth distribution policy. The most relevant one is an application that opens multiple TCP sockets. It would have a clear advantage over an application that only uses one socket since it will get more bandwidth. This could be addressed with an iteration of our policy which, rather than targeting inter-flow fairness, uses inter-app fairness. This could be achieved by looking at the process ID (PID) of each active flow and proportionally assigning bandwidth to the PID rather than the flows open by the process. Alternatively, more advanced scheduling mechanisms, such as round-robin or proportional fairness, could be integrated to address these limitations.

*A.5.3  BISCAY and other CCA.* Inspired by newer versions of BBR (v2 and v3) that integrate additional signals for enhanced end-to-end bandwidth estimation [98], BISCAY could also leverage congestion signals from other CCAs. From CUBIC and Reno, BISCAY could use packet loss to complement the end-to-end bandwidth determination method and make it more precise. Additionally, CUBIC/Reno could be used as a fallback CCA in case BISCAY detects that the bottleneck has shifted to the wired segment. Similarly, Explicit Congestion Notification (ECN) can complement the end-to-end metric. Other domain-specific CCAs such as DCTCP [25] rely on queue occupancy and packet reordering to determine when the bottleneck is reached. In particular, queue occupancy metrics could improve BISCAY's cellular bandwidth estimation by leveraging RLC-layer queue sizes obtained directly from the modem via OpenDiag. Finally, techniques from TCP Vegas and Compound TCP, BISCAY could give more importance to time-based metrics such as one-way delay (delay gradient) or jitter (used by real-time applications) to improve its bandwidth localization accuracy.

```
1   # This callback is triggered when:
2   # - Receive ACK
3   # - Timeout
4   # - Duplicate ACK
5   # - Explicit Congestion Notification (ECN)
6   biscay_congestion_window_callback():
7       # Transition between states based on network conditions
8       if state == STARTUP:
9           cwnd = SlowStart()
10          # Try to get bandwidth from KPIs
11          if getCellularBW() == OK:
12              # Swicth to Biscay when we get KPI-based bandwidth
13              state = BISCAY
14
15      # Biscay mode: the bottleneck is in the RAN
16      elif state == BISCAY:
17          # Get cellular bandwidth from KPIs (OpenDiag)
18          cellular_bw = getCellularBW()
19          # Get end-to-end bandwidth (BBR)
20          end_to_end_bw = getEndToEndBandiwdth(BBR)
21          # Get RTT
22          rtt = min(rtt, MeasureRTT(last_ack))
23          # Set CWND
24          cwnd = BandwidthDelayProduct(BWSplitPolicy(cellular_bw), rtt)
25          # Check bottleneck location
26          if cellular_bw > end_to_end_bw: # Bottleneck in the wired segment
27              # Trigger fallback mechanism
28              state = FALLBACK
29              # Set CWND using the wired-specific CCA selected (BBR)
30              cwnd = setCWNDfromFallbackCCA(BBR)
31
32      # Fallback mode: the bottleneck is in the wired segment
33      elif state == FALLBACK:
34          # Set CWND using the wired-specific CCA selected (BBR)
35          cwnd = setCWNDfromFallbackCCA(BBR)
36
37          # Get cellular bandwidth from KPIs (OpenDiag)
38          cellular_bw = getCellularBW()
39          # Get end-to-end bandwidth (BBR)
40          end_to_end_bw = getEndToEndBandiwdth(BBR)
41          # Check if the bottleneck has changed
42          if cellular_bw == end_to_end_bw: # Bottleneck in the cellular link
43              # Disable fallback mechanism
44              state = BISCAY
45              # Get RTT
46              rtt = min(rtt, MeasureRTT(last_ack))
47              # Set CWND
48              cwnd = BandwidthDelayProduct(BWSplitPolicy(cellular_bw), rtt)
49      return cwnd
50
51  # Function that splits the available bandwidth equally for all flows
52  BWSplitPolicy(bw):
53      # Get number of active flows (Including UDP)
54      num_flows = getNumberActiveFlows()
55      return bw/num_flows
```